\documentclass[preprint,floats,aps,nofootinbib,amssymb,11pt]{revtex4}

\usepackage{subfigure,epsfig}
\def\beq{\begin{equation}}
\def\eeq{\end{equation}}
\def\bea{\begin{eqnarray}}
\def\eea{\end{eqnarray}}

\def\bt{\tilde b}

\def\fbi{{\rm fb}^{-1}}

\def\pslash{\not{\hbox{\kern-4pt $p$}}}
\def\qslash{\not{\hbox{\kern-4pt $q$}}}
\def\lv{\not{\hbox{\kern-4pt $L$}}}
\def\lsim{\mathrel{\raise.3ex\hbox{$<$\kern-.75em\lower1ex\hbox{$\sim$}}}}
\def\gsim{\mathrel{\raise.3ex\hbox{$>$\kern-.75em\lower1ex\hbox{$\sim$}}}}
\def\ifmath#1{\relax\ifmmode #1\else $#1$\fi}

\usepackage{graphicx}
\usepackage{epstopdf}
\usepackage{bm}
\begin{document}
\draft
\renewcommand{\thefootnote}{\arabic{footnote}}

\begin{flushright}
MADPH-10-1560, NPAC-10-08 \\
\end{flushright}

\title{Testing  CP Violation in $ZZH$ Interactions at the LHC}
\bigskip
\author{Neil D. Christensen, Tao Han, and Yingchuan
Li\footnote{Email Address: neil@hep.wisc.edu,\ than@hep.wisc.edu,\  yli@physics.wisc.edu}}
\address{Department of Physics, University of Wisconsin, 1150 University Avenue,
Madison, WI 53706, U.S.A.}

\begin{abstract}
We study genuine CP-odd observables at the LHC to test the CP
property of the $ZZH$ interaction for a Higgs boson with mass below the threshold
to a pair of gauge bosons via the process $pp\to ZH \to \ell^{+}
\ell^{-}\  b\bar b$. We illustrate the analysis by including a CP-odd $ZZH$ coupling, and show how to
extract the CP asymmetries in the signal events. After selective kinematical cuts to
suppress the SM backgrounds plus an optimal Log-likelihood analysis, we find that, with a CP violating
coupling $\tilde{b}\simeq0.25$, a CP asymmetry may be established at a
3$\sigma$ (5$\sigma$) level with an integrated luminosity of 
%28 (46) $\fbi$ at the LHC.
about 30 (50) $\fbi$ at the LHC.
\end{abstract}

\maketitle

\vskip 0.1cm
\noindent

\section{Introduction}

The CERN Large Hardron Collider (LHC) will lead us to revolutionary
discoveries in particle physics. Observing the Higgs boson(s) is arguably the most
anticipated discovery at the LHC. Once a Higgs boson ($H$) is observed, a significant part
of the LHC program will be focused on determining the properties of it.

One of the most important aspects of the Higgs boson interactions will be its CP
property. Due to the clear need for new sources of CP violation beyond the Standard Model (SM)
to explain the baryon asymmetry of the Universe,
it is conceivable that an extended Higgs sector may be a primary source with or without spontaneous
CP violation \cite{CPv}.
There have been significant efforts in the literature to explore the
possibility to observe the effects of CP-violation in the Higgs sector at the 
LHC \cite{Chang:1992tu,Berge:2008wi,Godbole:2007cn}.
%,Gunion:1996xu .
If the Higgs boson is heavy enough to decay to multiple identifiable SM particles,
then it is quite feasible to construct  CP-odd observables.
Examples include $H\to W^+W^-,\ ZZ$ or $H\to t\bar t$ \cite{Chang:1992tu}.
%,Chang:1993jy}.
However, if the Higgs boson is light and only decays to light fermion pairs,
then there will not be enough information for constructing a CP-odd kinematical observable like the triple product
using the final state momenta of the Higgs decay.
The only exceptions are the subdominant channel $H \to \tau\tau$, subsequently decaying to
 at least two charged tracks plus missing neutrinos \cite{Berge:2008wi},
which suffers from a small branching fraction to the identifiable final state
and difficulties for event reconstruction. This may lead us to test the Yukawa coupling of $\tau \bar \tau H$.
For $HVV$ couplings, one may hope to access the clean decay channel $H\to ZZ^* \to 4\ell$ \cite{Godbole:2007cn},
if the Higgs mass is not too far below $2M_{Z}$.
Attempts have also been made to probe the CP property of the $t\bar t H$ coupling at hadron colliders via virtual Higgs effects in $t\bar{t}$ production \cite{Schmidt:1992et}, and via the direct associated production of $t\bar{t} H$ at the LHC \cite{Gunion:1996xu}.
However, it is very challenging to establish the signal of the $t\bar t H$ final state at the LHC experiments \cite{Aad:2009zza}.

In general, it is known to be quite challenging to construct genuine CP-odd
observables at the LHC. First, the initial state of the LHC, as a $pp$
collider, is not a CP eigenstate, in contrast to the neutrality property of an
$e^+e^-$ collider or a $p\bar p$ collider. As a result, one may have to seek for neutral
subprocesses (with $q\bar{q},\ gg$ initial states) or decays of CP
eigenstates at the LHC experiments.
Second, even with the initial states $q(p_1) \bar{q}(p_2)$ and
$g(p_1) g(p_2)$, they form a CP eigenstate only with special spin and color configurations, and in their
center-of-mass (c.m.) frame which is in general different from the lab
frame. These two reference frames are related by a longitudinal
boost that is unknown and different event by event.
Only with large statistics would one expect the difference to be averaged out.
Furthermore,  the symmetric proton beams at the LHC make it impossible to identify
the direction of a quark versus an anti-quark on an event-by-event basis,
which is often needed to specify a particle versus an anti-particle.

In a recent paper \cite{Han:2009ra}, some genuine CP-odd observables have been
proposed suitable for the LHC. One of them is the difference of the scalar
transverse momenta \cite{Schmidt:1992et,Dawson:1995wg}, or equivalently the transverse energies,
\begin{equation}
\label{Teven}
{\cal{O}}_1 \equiv p^+_T - p^-_T\quad  {\rm or}\quad  E^+_T - E^-_T,
\end{equation}
where $p^{}_T = \sqrt{p^2_x + p^2_y },\ \ E^{}_T = \sqrt{ p^2_T + m_f^2 }$,
with superscripts $\pm$ specifying the particle charges.  This observable is CP-odd but $\hat{T}$-even\footnote{This $\hat{T}$ is the naive time-reversal transformation which only flips the sign of momenta and spins, without changing the initial or final states.} and thus generated from CP-violation associated with the absorptive part of the amplitude \cite{Valencia:1994zi}, thus relying on the existence of
an additional CP-conserving strong phase $\sin\delta$. The other one is a modified triple product 
\begin{equation}
\label{Todd}
{\cal{O}}_2 \equiv (\hat p_f \times \hat p_{\bar f} ) \cdot \hat{z}\
{\rm sgn}( (\vec p_f - \vec p_{\bar f})\cdot \hat{z}),
\end{equation}
where $\vec{p}_{f,\bar{f}}$ are the 3-momenta of the particle $f$ and anti-particle $\bar f$, $\hat{p}_{f,\bar{f}}$ are their unit vectors, 
and $\hat{z}$ is the beam direction. 
This observable is CP-odd and $\hat{T}$-odd, and may be generated from the dispersive part of the amplitude \cite{Valencia:1994zi}.
Both variables (\ref{Teven}) and (\ref{Todd}) are independent of the choice of a quark momentum direction and are longitudinally boost
invariant, thus adequate for LHC experiments. 
 
\begin{figure}
\begin{center}
\includegraphics[scale=0.60]{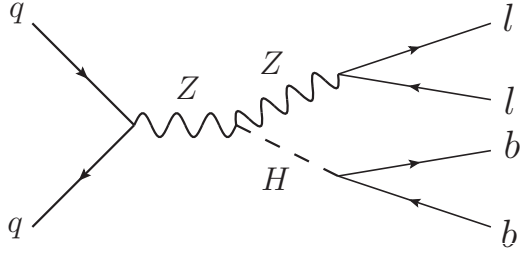}
\end{center}
\caption{ \label{signalDiagram}Feymann diagram of signal process at the parton level $q \bar{q} \rightarrow ZH \to \ell^{+}\ell^{-}\ b\bar b$.}
\end{figure}

In this article, we address the feasibility of discovering CP violation in $ZZH$ interactions using these observables.  We propose to analyze the process
\begin{equation}
p p \to Z H \to \ell^+ \ell^-\ b\bar b ~~{\rm with}~ \ell = e,\mu
\label{eq:fs}
\end{equation}
where the $Z$ decays leptonically and the Higgs to $b\bar{b}$.  The signal diagram is illustrated in Figure \ref{signalDiagram}.
The most general Lorentz structure of the Higgs and Z boson coupling, $ H Z^{\mu}(k_1)Z^{\nu}(k_2)$, is
\begin{equation}
\Gamma^{\mu\nu}(k_1,k_2) = i \frac{2}{v} \left[ a M^2_Z g^{\mu\nu}
+ b ~(k^{\mu}_1 k^{\nu}_2 -  k_1 \cdot k_2 g^{\mu\nu}) + \tilde{b}~
\epsilon^{\mu\nu\rho\sigma} k_{1\rho} k_{2\sigma}\right],
\label{eq:operator}
\end{equation}
where $v=(\sqrt{2}G_F)^{-1/2}$ is the vacuum expectation value of the Higgs field, and the $Z$ boson 4-momenta are both incoming. The $a$ and $b$ terms are CP-even while $\tilde{b}$ is CP-odd. The simultaneous presence of both $a$
(or $b$) and $\tilde{b}$ in this vertex would generate CP-violation.
In the SM at tree level, $a=1$ and $b=\tilde{b}=0$. For a given theory of the SM or beyond, radiative corrections can generate nonzero contributions to $a$, $b$ and $\tilde b$,
and thus each of these are energy-dependent form factors in general \cite{Cao:2009ah}.
In effective field theory language,
$b$ and $\tilde{b}$ can be obtained from gauge invariant dimension-6 operators such as
\begin{equation}
{c\over \Lambda^{2 }} H^{\dagger}H V_{\mu\nu}V^{\mu\nu}\quad \mbox{and} \quad
{\tilde c\over \Lambda^{2}} H^{\dagger}H \tilde{V}_{\mu\nu}V^{\mu\nu}
\end{equation}
respectively \cite{Hagiwara:1993ck}.  Depending on the nature of the underlying theory, the effective couplings
$c,\ \tilde c$ could be of order unity for a weakly coupled theory, or of order $(4\pi)^{2}$ 
in a strongly coupled theory. 
If the theory is valid up to a scale $\Lambda \sim 4\pi v$, we expect that $b$ and $\tilde{b}$ should be of the order
$1/(4\pi)^{2}$ to 1.
%
%Since the CP property is independent of the momentum dependence in form factors,
For our phenomenological studies, we will simply take them as constants with some optimistic values, and will focus on the complex CP-violating parameter $\tilde{b}$ which accommodates both dispersive (real $\tilde{b}$) and absorptive (imaginary $\tilde{b}$) CP violation.

The rest of this paper is organized as follows: In Sec.~\ref{sec:asymmetry}, we concentrate on the signal and show that an asymmetry appears in the observables  ${\cal{O}}_{1,2}$ with the above theoretical parameterizations.  In Sec.~\ref{sec:reach}, we consider the backgrounds 
to this process at the LHC, apply judicial kinematic cuts and determine the integrated luminosity necessary to observe these asymmetries.
In Sec.~\ref{sec:discussion}, we discuss our results further.  Finally, in Sec.~\ref{sec:conclusion}, we conclude.

%
%The rest of work is organized as follows.
%In Section II, we set up the formalism by introducing CP-odd $ZZH$ coupling.
%In Section III, we quantify the SM backgrounds and present the sensitivity reach
%for the CP asymmetry. We conclude in Section IV.

%\section{CP-odd observable in $ZH\to \ell^{+}\ell^{-}\ b\bar b$ }
%\label{sec:general}

% ------------------------------------------------------------------------------------------------------------------

\section{CP asymmetries in $pp\rightarrow ZH \rightarrow \ell^+ \ell^- b \bar{b} $ at the LHC}
\label{sec:asymmetry}

The CP asymmetries manifest themselves in the distributions of the CP-odd observables ${\cal{O}}_{1,2}$
as introduced in the previous section. We find it convenient to introduce
\begin{equation}
\Delta E_T \equiv  E^+_T - E^-_T\quad\mbox{and}\quad
\phi_{ll} \equiv {\rm sgn}((\vec{\ell}^+ - \vec{\ell}^-)\cdot \hat{z})\ {\sin}^{-1} (\hat{\ell}^+ \times \hat{\ell}^- \cdot \hat{z}),
\end{equation}
% d\sigma/d\Delta E_T$ and $d\sigma/d\Phi$, where
% $ is the transverse energy difference of two leptons,
where $\phi_{ll}$ is the azimuthal angle between the two lepton planes 
$\vec{\ell}^+ - \hat{z}$ and $\vec{\ell}^- -\hat{z}$, 
multiplied by the sign of their logitudinal momenta difference.
%the two  multiplied by the sign of their longitudinal momenta difference
%The Feymann diagram of $ZH$ production at parton level is shown in Figure \ref{signalDiagram}.
Our study of the signal is based on a parton-level Monte Carlo simulation
of $pp\rightarrow ZH \rightarrow \ell^+ \ell^- b \bar{b}$, incorporating the full spin correlations
from the production to the decay.
We use the CTEQ6.1L parton distribution functions \cite{Pumplin:2002vw}.
We simulate the LHC $pp$ collisions at the c.m.~energy $\sqrt{s}=14$ TeV.
At lower c.m.~energies, we would not expect any qualitative change for our asymmetry discussions,
although the signal cross section and SM backgrounds will be different.

\begin{figure}[tb]
\centerline{ \epsfig{file=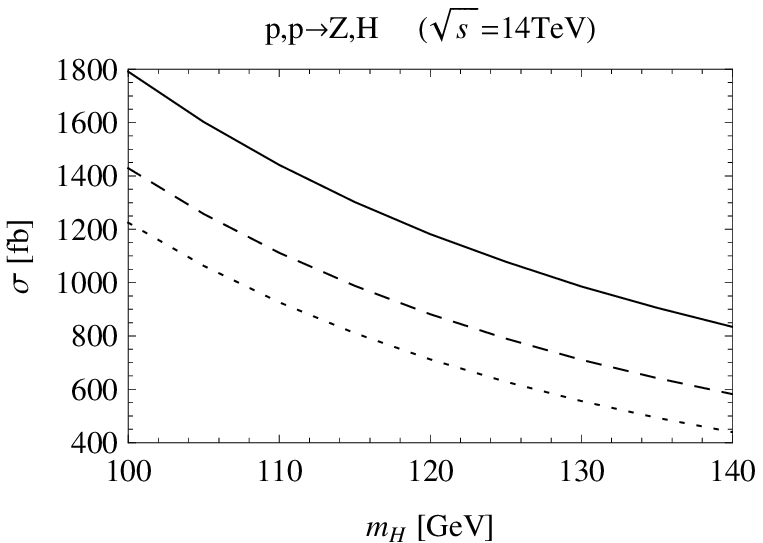,width=7.5cm} ~~ \epsfig{file=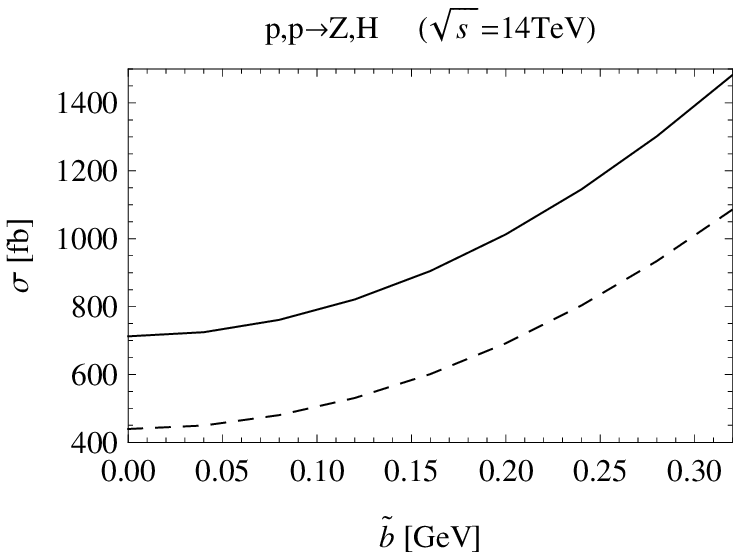,width=7.5cm}}
\caption{ \label{fig:tot}Total production cross-section for $pp\rightarrow ZH $  at the LHC. 
The left panel contains the cross-section versus $m_{H}$ for $\bt=$0 (dotted), 0.15 (dashed) and 0.25 (solid).  The right panel contains the cross-section
versus $\bt$ for $m_{H}=120$ (solid) and 140 (dashed) GeV. 
}
\end{figure}

We first present the total cross-sections for the signal versus Higgs mass for 
$\tilde b=0, 0.15$ and $0.25$ in the left panel of Fig.~\ref{fig:tot}.
 We see that, for $m_H=120$ GeV, the SM cross section 
is of order 700 fb and $\tilde{b}$ enhances the cross-section 
by $23\%$ ($65\%$) for $\tilde{b}=0.15$ ($\tilde{b}=0.25$).  In the right panel of Fig.~\ref{fig:tot}, 
we show the signal cross section as a function of $\tilde{b}$ for $m_H=120$ and $140$ GeV. There is no decay branching fraction included
in these figures.   We use the illustrative value $m_H=120$ GeV in the remainder of this paper.

\begin{figure}[tb]
\centerline{ \epsfig{file=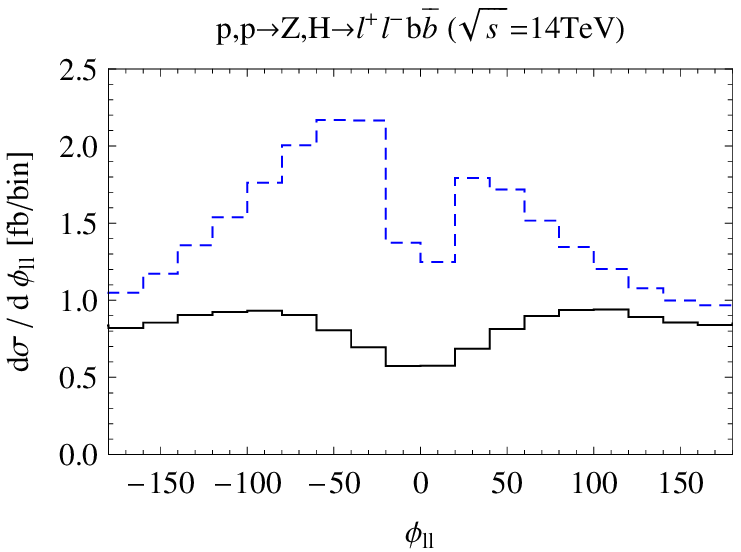,width=7.5cm} ~~ \epsfig{file=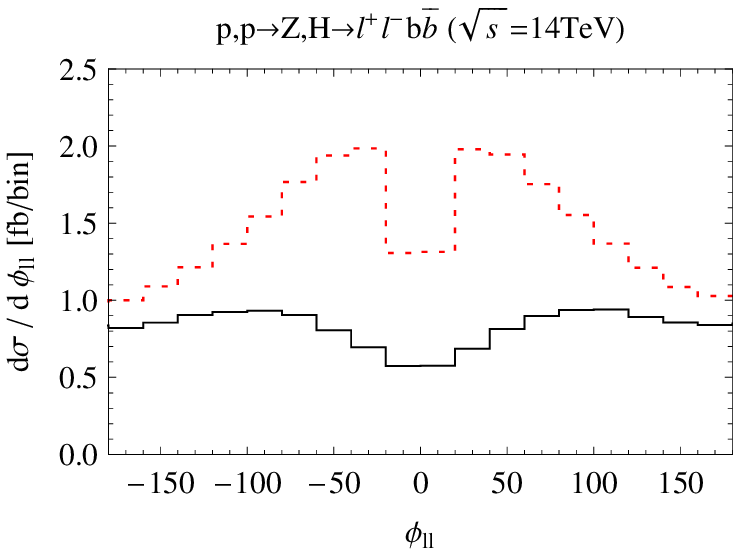,width=7.5cm}}
\caption{\label{fig:signalasy1}$d\sigma/d\phi_{ll}$ distributions.  The solid curve in both plots is for the SM expectation.  The dashed curve on the left is for $\tilde{b}=0.25$ while the dotted curve on the right is for $\tilde{b}=0.25i$.  The bin size is $18^\circ$.}
\end{figure}

\begin{figure}[tb]
\centerline{ \epsfig{file=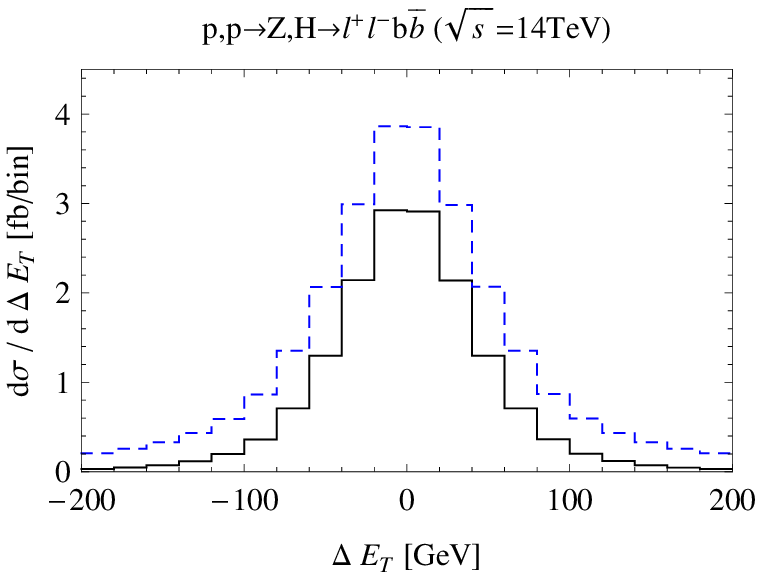,width=7.5cm} ~~
\epsfig{file=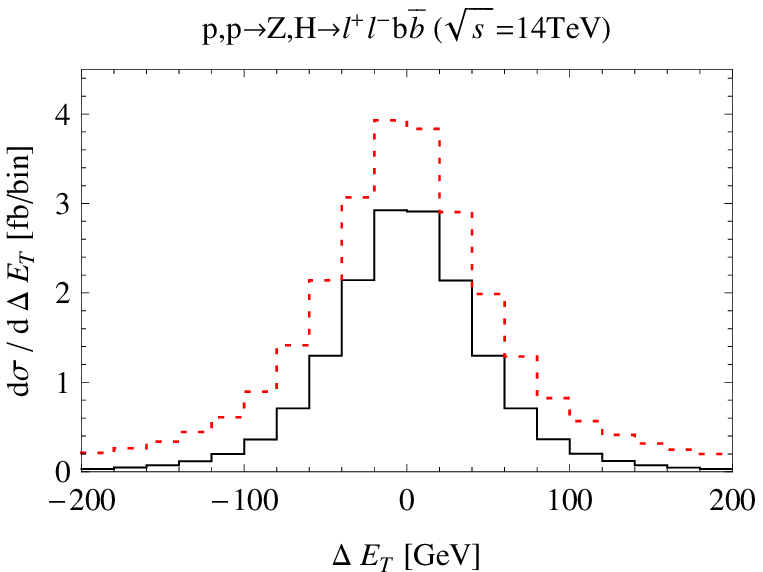,width=7.5cm}  }
\caption{\label{fig:signalasy2}$d\sigma/\Delta E_T$ distributions.  The solid curve is for the SM expectation in both plots.  The dashed curve on the left is for $\tilde{b}=0.25$, while the dotted curve on the right is for $\tilde{b}=0.25i$. The bin size is 10 GeV.}
\end{figure}

To simulate geometrical coverage of the detector and for the purpose of triggering,
we impose the following basic cuts of transverse momentum, rapidity, and separation 
on the jets and leptons:
\begin{equation}
\label{eq:cuts}
p_{T_{j,l}} > 20 ~{\rm GeV},
~~|\eta_{j,l}| < 2.5
~~\mbox{and} ~~ 
\Delta R_{jj,jl} > 0.4.
\end{equation}
With these acceptance cuts, we show the distribution of $d\sigma/d\phi_{ll}$ in Fig.~\ref{fig:signalasy1} 
for the case $\tilde{b}=0.25$ (left panel), and the case with $\tilde{b}=0.25 i$ (right panel) by the dashed histograms. The SM result is included for comparison by the solid curves in both panels. 
We see that the CP asymmetry in $d\sigma/d\phi_{ll}$ is only induced by CP-violation in the dispersive amplitude as in the left panel
with real $\tilde{b}$, but not in the absorptive amplitude as in the right panel with imaginary $\tilde b$. 
In contrast, the distribution of  $d\sigma/d\Delta E_T$ is shown in Fig.~\ref{fig:signalasy2}
 with the same parameter choice, and 
the CP asymmetry in this variable is only induced by CP-violation in the absorptive amplitude in the right panel with imaginary $\tilde{b}$, 
but not in the dispersive amplitude in the left panel with real $\tilde b$. 
 %
%Moreover, since a CP-conserving strong phase could also be induced from the width of the $Z$, the CP asymmetry in $d\sigma/d\Delta E_T$ is, in principle, also generated from the interference of Re$(\tilde{b})$ with the width. However, that effect is much smaller and appears not to be observable in practice.

\begin{figure}[tb]
\centerline{ \epsfig{file=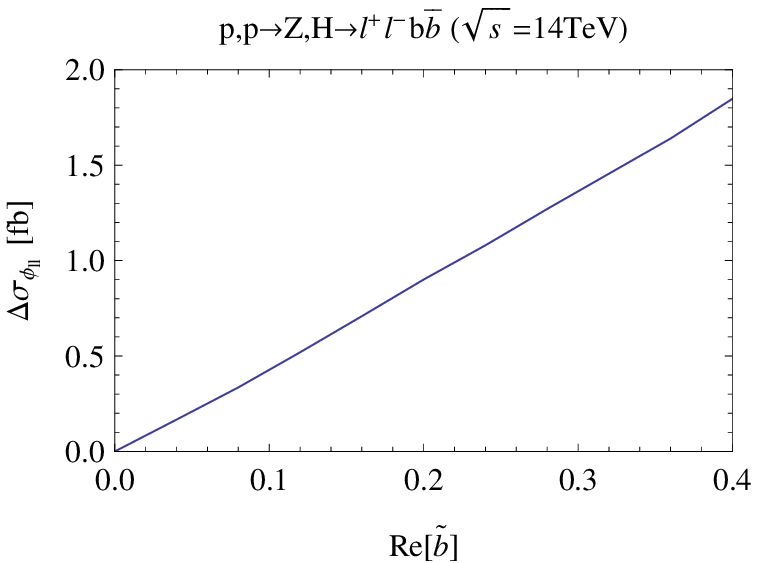,width=8.2cm} 
\epsfig{file=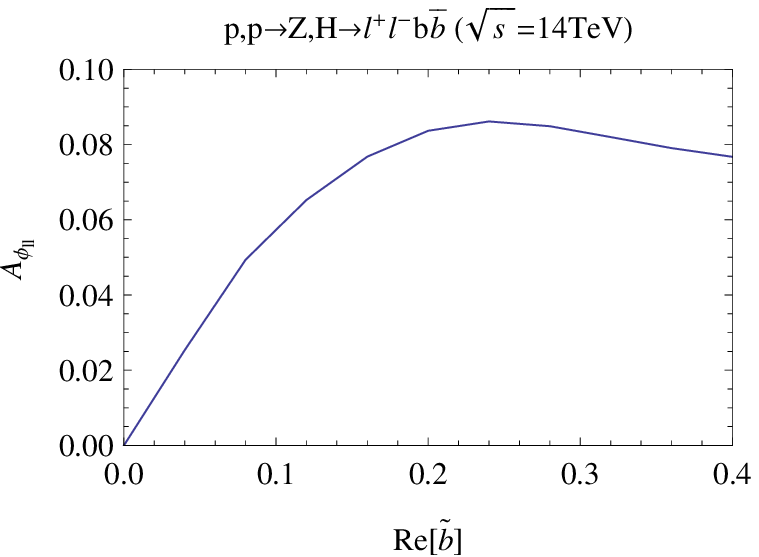,width=8.2cm}  } 
\caption{\label{fig:phibtilde}
The cross-section difference $\Delta \sigma_{\phi_{ll}}$ 
and the corresponding asymmetry $A_{\phi_{ll}}$ as functions of Re($\bt$).}
\end{figure}

\begin{figure}[tb]
\centerline{ \epsfig{file=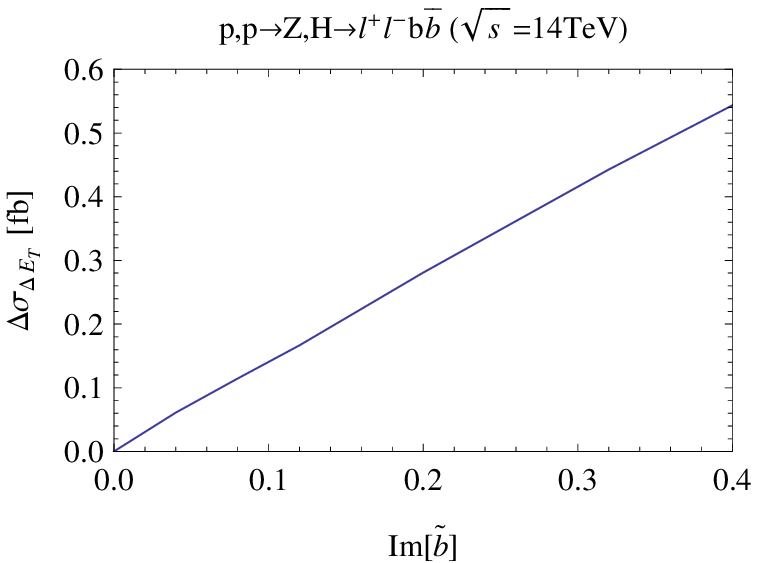,width=8.2cm} 
\epsfig{file=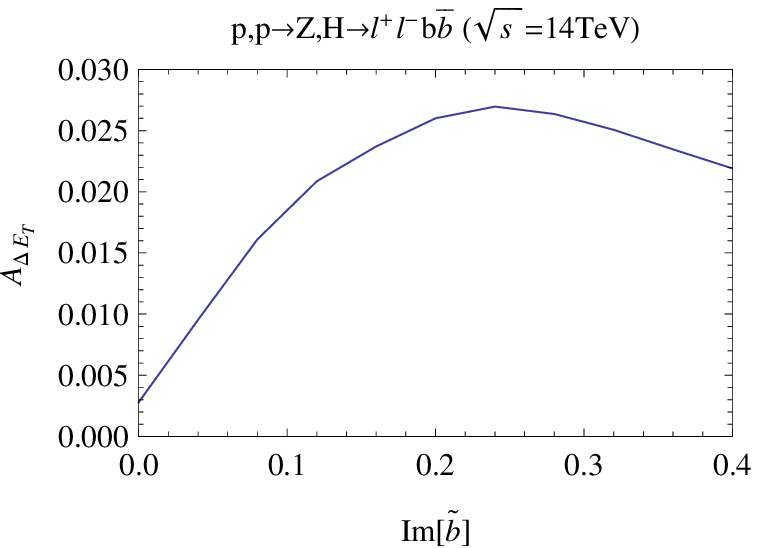,width=8.2cm}  }
\caption{\label{fig:deltaETbtilde} The cross-section difference $\Delta \sigma_{\Delta E_T}$ 
and the corresponding asymmetry $A_{\Delta E_T}$ as functions of Im($\bt$).}
\end{figure}

The asymmetrical distributions in $\phi_{ll}$ and $\Delta E_T$ induced by CP violation leads to nonzero values of the cross-section differences
\begin{equation} 
\Delta \sigma_{\phi_{ll}} \equiv \sigma_{\phi_{ll}<0} - \sigma_{\phi_{ll}>0} 
~~\mbox{and}~~ 
\Delta \sigma^{}_{\Delta E_T} \equiv \sigma^{}_{\Delta E_T<0} - \sigma^{}_{\Delta E_T>0},
\end{equation}
and in the corresponding asymmetries which are conventionally defined as 
\begin{equation}
A_{\phi_{ll}} \equiv \frac{\sigma_{\phi_{ll}<0} -
\sigma_{\phi_{ll}>0}}{\sigma_{\phi_{ll}<0} +
\sigma_{\phi_{ll}>0}} 
~~\mbox{and}~~ 
A^{}_{\Delta E_T} \equiv \frac{\sigma^{}_{\Delta E_T<0} - \sigma^{}_{\Delta E_T>0}}{\sigma^{}_{\Delta E_T<0} + \sigma^{}_{\Delta E_T>0}}.
\end{equation}
We show the cross-section difference $\Delta \sigma_{\phi_{ll}}$ ($\Delta \sigma_{\Delta E_T}$)
and corresponding asymmetries $A_{\phi_{ll}}$ ($A_{\Delta E_T}$) in Fig.~\ref{fig:phibtilde} 
(\ref{fig:deltaETbtilde})
 for a range of $\bt$ values. The cross-section differences may be sizable and could reach 
about 1 fb for  $\Delta \sigma_{\phi_{ll}}$ and 0.3 fb for $\Delta \sigma_{\Delta E_T}$ with $|\tilde b|\sim 0.25$.
The asymmetry in $A_{\phi_{ll}}$ is typically larger than that in $A_{\Delta E_T}$. 
We see that the cross-section difference follows a linear-dependence on $|\bt |$, while the asymmetries 
reach their maxima around Re$(\bt) = 0.25$ and Im$(\bt) = 0.25$ for $A_{\phi_{ll}}$  and $A_{\Delta E_T}$, respectively. 
This is an indication that the higher order terms in $\bt$ become significant, and thus care needs to be taken 
for larger values of $|\bt |$ when interpreting the results of the asymmetries. 
%{\bf (it's kind of surprising that cross section diff is dominated by the linear term (quadratic terms cancel); while the quadratic terms
%in asymmetry seem to take over at $\bt >$0.25). Do you agree? YC: I think this is as expected.  NC: The $\tilde{b}^2$ term is symmetric and %cancels in difference and so does not affect $\Delta\sigma$.  However, it is present in the sum $\sigma_T$ in the denominator of $A$.  %Furthermore, it grows with twice the power of momentum that the linear term does, so it rapidly dominates over the linear term.}
%

% ----------------------------------------------------------

\section{Observability of the CP asymmetries at the LHC}
\label{sec:reach}
\begin{figure}
\centerline{\epsfig{file=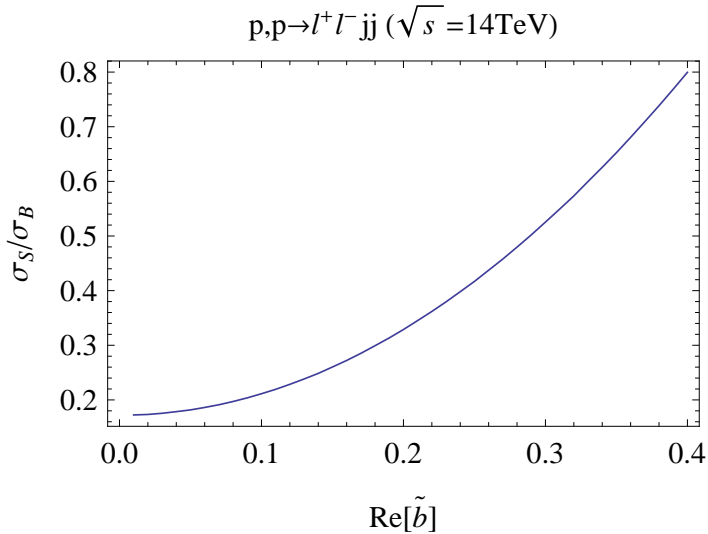,width=10.0cm}}
\caption{\label{SoB}Signal over background distribution.}
\end{figure}

The signals we are searching for are the $ZH$ events as in Eq.~(\ref{eq:fs}).
We specify these events with two jets, with at least one $b$-tagged, and two opposite-sign leptons of the same flavor, either electron or muon.  Contributions to the background mainly come from three sources:
\begin{eqnarray}
pp \rightarrow {\rm QCD} \ b\bar{b}\ + \ell^+ \ell^- , \quad
pp \to  t\bar{t} \rightarrow b\bar{b}\ \ell^+ \ell^-\ \nu\bar{\nu} , \quad
pp \rightarrow {\rm QCD}\ jj\ + \ell^+ \ell^- .
\label{eq:bckgrnd}
\end{eqnarray}
These backgrounds are calculated using the CalcHEP package \cite{Pukhov:1999gg,Pukhov:2004ca}.  The full spin correlation have been kept for all $pp\rightarrow jj \ell^+ \ell^-$ where the full $2\rightarrow4$ processes are calculated.  In the case of the $t\bar{t}$ background, $pp\rightarrow t\bar{t}$ events are generated and then decayed, ignoring spin correlation.  Since this process is subdominant, the error resulting from loss of spin correlation should not be large.
%
%  The dominant background is from the process $pp\rightarrow b\bar{b}l^+l^-$ where at least one jet is b tagged.  Another important background came from the process $pp\rightarrow t\bar{t}\rightarrow b\bar{b}l^+l^-\nu\bar{\nu}$ where at least one jet is b tagged and the missing energy satisfies either $p_T<20GeV$ or $\left|\eta\right|>2.5$.
%
The processes $pp\rightarrow jj \ell^+ \ell^-$, where $j$ denotes a jet from a light quark ($u,d,s$ or $c$) or gluon, yield 
large rates due to QCD. Demanding at least one tagged $b$-jet substantially reduces this background. 
The tagging and mistagging rates we use are taken from Figure 12.30 of the CMS TDR \cite{CMS-TDR}, 
and are listed here in Table \ref{efficiencies}. 
\begin{table}[tb]
\begin{tabular}{|l|c|}
\multicolumn{2}{c}{Jet tagging/mistagging rates}\\
\hline
b-quark jet & $0.4$\\
\hline
c-quark jet & $0.03$\\
\hline
Gluon jet & $0.006$\\
\hline
light jet  & $0.001$\\
\hline
\end{tabular}
\caption{\label{efficiencies}The probability that a jet will be tagged as a b-quark jet.  Taken from the CMS TDR Figure 12.30.}
\end{table}
The mistagging rates are correlated with the tagging efficiency.
We have taken a rather low value for the $b$-tagging efficiency (0.4), 
in order to substantially lower mistagging rates and thus to significantly reduce the backgrounds coming from non-$b$-quark jets.
Our signal events should not have much missing energy.  We can, therefore, significantly reduce the $t\bar{t}$ backgrounds by demanding that the mssing energy in these events satisfies 
\begin{equation}
\slash\hspace{-0.075in}p_T < 20~\mbox{GeV} 
~~\mbox{or}~~
 \left|\slash\hspace{-0.075in}\eta\right|>2.5
 \label{eq:missingCuts}.
\end{equation}
We envision the search for this asymmetry occurring after the Higgs has already been discovered and its mass is known.  We have, therefore, cut the invariant mass of the jets to be near the Higgs mass.  Further, the $b$ jets from the Higgs decay tend to be energetic, typically around the Jacobean peak near $M_H/2$.  For this reason, we apply a higher transverse momentum cut on the jets.  Our final cuts for this process are
\begin{equation}
p_{T_j} > 50 ~{\rm GeV}, ~~
|M_{jj}-M_H| < 10 ~{\rm GeV}
~~\mbox{and}~~
|M_{ll}-M_Z| < 15 ~{\rm GeV}
\label{eq:finalCuts}.
\end{equation}
further suppressing the background.  With these cuts, the signal over background can be seen in Figure \ref{SoB}.  We see that the background rate is much greater than the signal for small $\tilde{b}$.

The angular distributions of the background processes (see Eq.~(\ref{eq:bckgrnd}))
\begin{figure}
\centerline{\epsfig{file=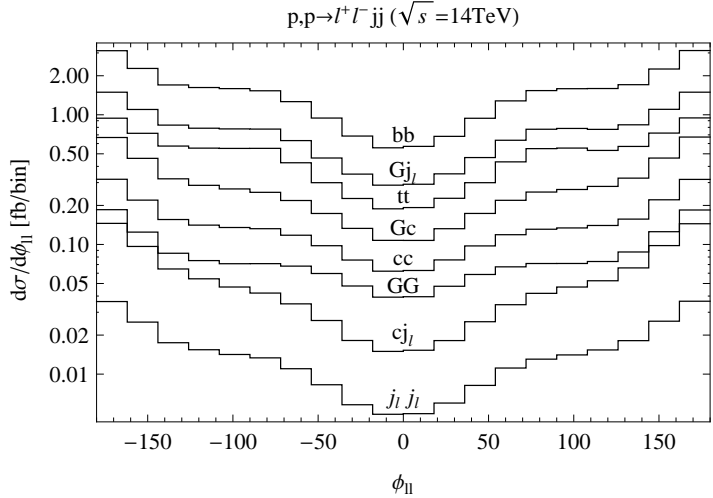,width=10.0cm}}
\caption{ \label{backgrounds}Contributions to the background as in Eq.~(\ref{eq:bckgrnd}).  
Many channels are presented separately as labeled. 
Each background is added to the previous one so that the top curve gives the total background.
The full set of cuts given in Eqs.~(\ref{eq:cuts}), (\ref{eq:missingCuts}) and (\ref{eq:finalCuts}) 
is applied as well as $b$-tagging on at least one jet.
%
%The line labeled `bb' is the dominant background and corresponds to $pp\rightarrow b\bar{b}l^+l^-$ where at least one jet was b-tagged.  The line labeled `tt' corresponds to $pp\rightarrow t\bar{t}\rightarrow b\bar{b}l^+l^-\nu\bar{\nu}$ where at least one jet is b-tagged and the missing energy satisfies either $p_T<20GeV$ or $\left|\eta\right|>2.5$.   All other lines are for $pp\rightarrow jjl^+l^-$ where at least one jet was mistagged as a b-quark jet.  `G' stands for a gluon jet, `c' represents a c-quark jet and `$j_l$' is a u-quark, d-quark or s-quark jet.  The b-quark jet tagging and mistagging rates were taken from the CMS TDR Figure 12.30 and are given in Table \ref{efficiencies}.  Full spin correlation was kept for all processes except the process $pp\rightarrow t\bar{t}\rightarrow b\bar{b}l^+l^-\nu\bar{\nu}$.  Since this process was subdominant, it was felt that the error induced by loss of spin correlation was not significant. 
%
 The bin width is $18^\circ$.  
}
\end{figure}
are shown in Fig.~\ref{backgrounds}. Many different final state partons have been separately presented for clarity. 
The leading background comes from QCD $b\bar{b}\ \ell^+ \ell^-$, where at least one jet is $b$-tagged.
The sub-leading background is from $qg \to qg \ell^{+}\ell^{-}$ (labelled as $Gj_{l}$), and the next one is due to $t\bar t$.
%
%  The line labeled `tt' corresponds to $pp\rightarrow t\bar{t}\rightarrow b\bar{b}l^+l^-\nu\bar{\nu}$ where at least one jet is b-tagged and the missing energy satisfies either $p_T<20GeV$ or $\left|\eta\right|>2.5$.   All other lines are for $pp\rightarrow jjl^+l^-$ where at least one jet was mistagged as a b-quark jet.  `G' stands for a gluon jet, `c' represents a c-quark jet and `$j_l$' is a u-quark, d-quark or s-quark jet.  The b-quark jet tagging and mistagging rates were taken from the CMS TDR Figure 12.30 and are given in Table \ref{efficiencies}.  Full spin correlation was kept for all processes except the process $pp\rightarrow t\bar{t}\rightarrow b\bar{b}l^+l^-\nu\bar{\nu}$.  Since this process was subdominant, it was felt that the error induced by loss of spin correlation was not significant.  The bin width is $18^\circ$. 
%
The full set of cuts given in Eqs.~(\ref{eq:cuts}), (\ref{eq:missingCuts}) and (\ref{eq:finalCuts}) 
are applied along with $b$-tagging of at-least one jet.
Each background is added to the previous one so that the top curve gives the total background. The shapes of the background distributions are qualitatively different from that of the signal, while the total rate is much larger than the signal for small $\tilde{b}$ (see Figure \ref{SoB}.)

\begin{figure}[tb]
\centerline{ \epsfig{file=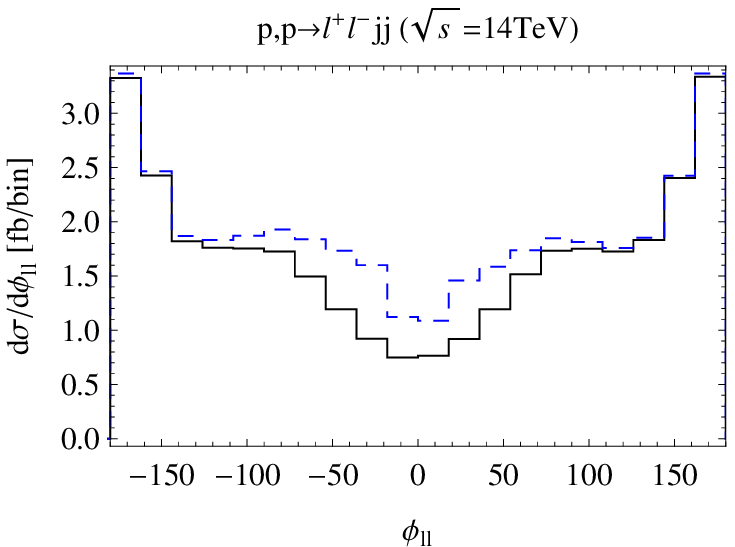,width=7.5cm} ~~ \epsfig{file=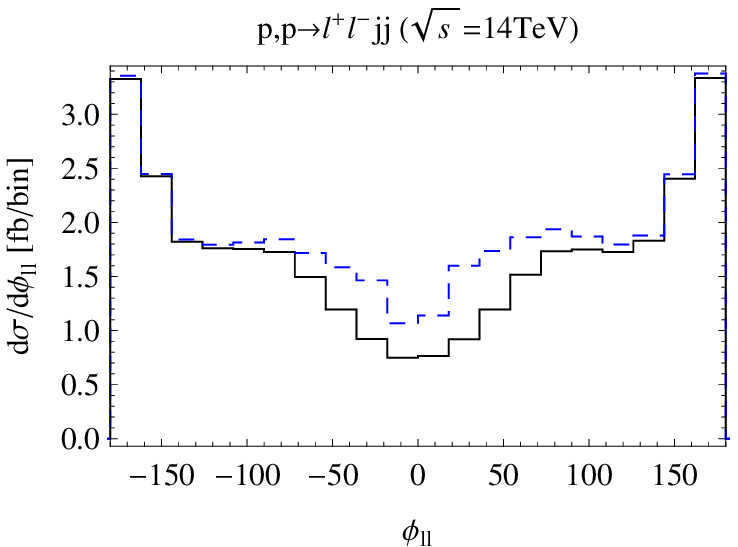,width=7.5cm} }
\caption{ \label{signal}Differential cross section for background plus signal.  The solid line in both plots is for $\tilde{b}=0$ (SM).  The dashed line on the left is for $\tilde{b}=0.25$ while the dashed line on the right is for $\tilde{b}=-0.25$.  
All backgrounds are included as in Figure \ref{backgrounds}.  The bin width is $18^\circ$.}
\end{figure}

\begin{figure}[tb]
\centerline{ \epsfig{file=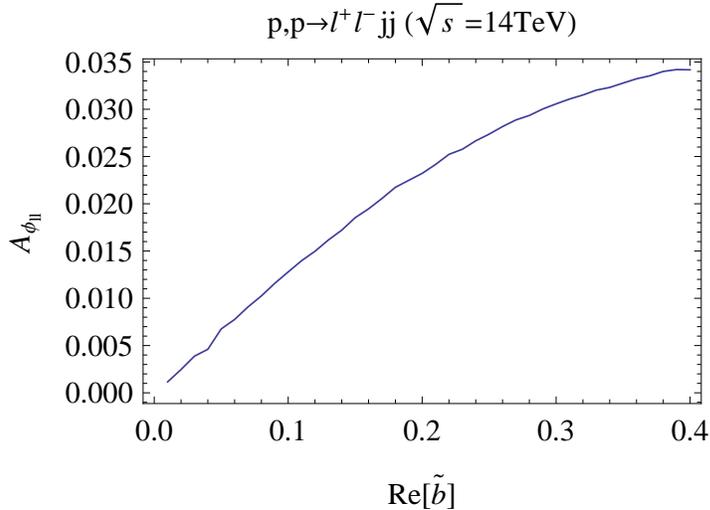,width=10.0cm} }
\caption{\label{asymmetry}Asymmetry $A_{\phi_{ll}}$ as a function of $\tilde{b}$.  
%The 14 inner bins (from $-126^\circ$ to $126^\circ$) were used to calculate the asymmetry.  
All backgrounds were included.  The inner range of Eq.~(\ref{eq:inner}) was used to calculate the asymmetry. }
\end{figure}

Example distributions are shown in the left panel and the right panel of Fig.~\ref{signal}
for $\tilde{b}=0.25$ and $\tilde{b}=-0.25$, 
respectively. The solid curve in each plot is the SM expectation, and the dashed curve corresponds to the sum of the signal and backgrounds.
It is reassuring to see that the asymmety changes sign with a change of the sign of $\bt$, which indicates a dominance of the linear contribution of $\tilde{b}$ to the asymmetry and thus justifies the leading effective operator approximation. 
As found earlier, the signal asymmetry occurs not far from $\phi_{ll}=0$, while the background tends to be larger near
$\phi_{ll}=\pm \pi$. This thus motivates us to evaluate the asymmetry $A_{\phi_{ll}}$ in a  restricted region
\begin{equation}
-126^{\circ} < \phi_{ll} < 126^{\circ}. 
\label{eq:inner}
\end{equation}
The resulting asymmetry is shown in Fig.~\ref{asymmetry},
where all backgrounds have been taken into
account.  We see that the measured asymmetry could reach percentage level.

\begin{figure}
\centerline{\epsfig{file=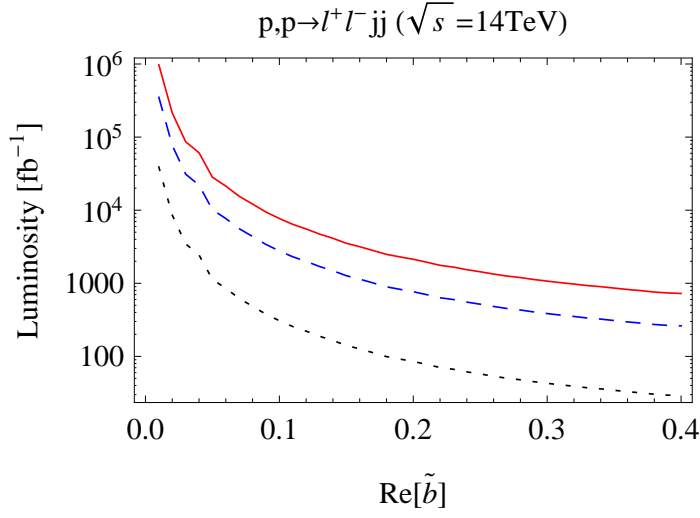,width=10.0cm}}
\caption{\label{fig:Gauss}The integrated luminosity required to measure the asymmetry $A_{\phi_{ll}}$ 
at 1$\sigma$ (dotted line), 3$\sigma$ (dashed line) and 5$\sigma$ (solid line) using a Gaussian estimate of the significance.}
\end{figure}

A straightforward Gaussian estimate of the significance is given by 
\begin{equation}
S = \frac{ | A | }{\delta A} \simeq \frac{ | \Delta\sigma_{\phi_{ll}} |}{\sqrt{\sigma_T}} \sqrt{L}
\end{equation}
where $\sigma_T=\sigma_{\phi_{ll}<0} + \sigma_{\phi_{ll}>0} $ is the total cross section and $L$ is the integrated luminosity.   Inverting this gives the integrated luminosity required to discover this asymmetry at an $S\sigma$ significance
\begin{equation}
L = S^2\frac{\sigma_T}{\left(\Delta\sigma_{\phi_{ll}}\right)^2} .
\end{equation}
The integrated luminosity to measure this asymmetry at 1$\sigma$ (dotted line), 3$\sigma$ (dashed line) and 5$\sigma$ (solid line) level is given in Figure \ref{fig:Gauss}.
For instance, for $\tilde{b}=0.25$, one would need an  
integrated luminosity of about 60, 520 or 1440 fb$^{-1}$ to reach 1, 3 or 5$\sigma$ sensitivity, respectively. 

\begin{figure}[tb]
\centerline{\epsfig{file=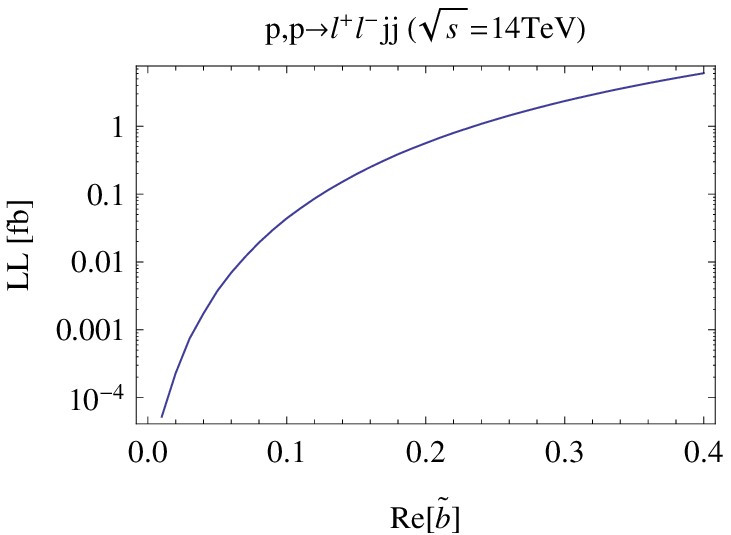,width=8.4cm}~\epsfig{file=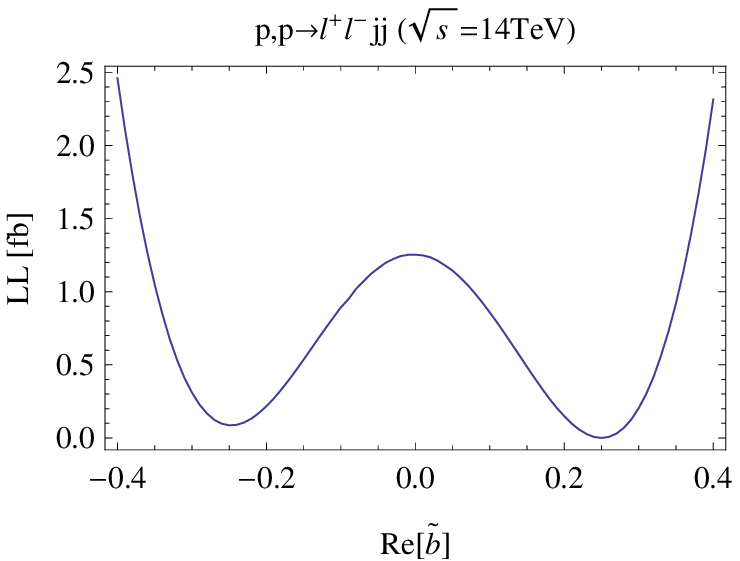,width=8cm}} 
\caption{\label{LLvsBt}Log likelihood in units of fb as a function of $\tilde{b}$.   The 14 inner bins (from $-126^\circ$ to $126^\circ$) were used to construct these plots.  The left panel shows the LL with respect to the SM for a range of $\tilde{b}$.  The right panel shows the LL of $\tilde{b}=0.25$ with respect to a range of $\tilde{b}$. 
% for $\tilde{b}=0.25$ with respect to other values of $\tilde{b}$.  A value of $\tilde{b}=0.25$ is significantly different than the SM while less significantly different than $-\tilde{b}$.  This means that it will be easier to measure the absolute value of $\tilde{b}$ than its sign.  We estimate the integrated luminosity required to measure each in Figure \ref{fig:lLL}.    The 14 inner bins (from $-126^\circ$ to $126^\circ$) were used to calculate the log likelihood. 
% For $\tilde{b}<0.6$ the LL at $-\tilde{b}$ is larger than it is at $\tilde{b}=0$.  For $\tilde{b}>0.6$, the LL at $\tilde{b}=0$ is larger than it is at $-\tilde{b}$.  In each case, we take the more conservative estimate.  For $\tilde{b}<0.6$, we take the LL with respect to the SM ($\tilde{b}=0$) while for $\tilde{b}>0.6$, we take the LL with respect to $-\tilde{b}$.  This ensures that we are always able to distinguish $\tilde{b}$ from both the SM ($\tilde{b}=0$) and from $-\tilde{b}$. 
}
\end{figure}

The sensitivity can be significantly improved if we consider a $\chi^2$ distribution of the binned data.  The ``log likelihood'' (LL) is defined to be
\begin{equation}
LL  = 2\sum_i\left[n_i\mbox{ln}\left(\frac{n_i}{\nu_i}\right)+\nu_i-n_i\right] ,
\end{equation}
where $n_i$ is the number of events observed in bin $i$ where $\nu_i$ are expected.  The LL determines the likelihood that the expected number of events in each bin could fluctuate up to mimic what is observed.  The larger the LL, the less likely it is a fluctuation.  
We plot the LL as a function of $\tilde{b}$ normalized to units of fb in Fig.~\ref{LLvsBt}(a) where we have taken the expectation ($\nu_i$) to be that of the SM ($a=1$, $b=\tilde{b}=0$). 
Once a signal observation is indicated, it would be  important  to address whether one is able to determine the sign of $\tilde{b}$, which not only
checks the consistency of the leading operator approximation, but would also give a hint for the underlying physics responsible for the
CP violating interactions. 
We plot the LL as a function of $\tilde{b}$ normalized to units of fb in Fig.~\ref{LLvsBt}(b) but  take the observation ($n_i$) to be that 
for  $a=1$, $b=0$ and $\tilde{b}=0.25$ while the expectation ($\nu_i$) is given by the value of $\tilde{b}$ along the horizontal axis. 
It is interesting to see that the results for $\tilde{b}=\pm 0.25$ are indeed more similar and it may take more data to distinguish
the sign ambiguity for a large value of $\tilde{b}$.

In the limit of large $n_i$, the LL approaches a $\chi^2$ distribution.  To determine the value of the LL corresponding with the Gaussian 1, 3 and 5$\sigma$ deviations, we equate the probabilities of fluctuating up to that point or higher as in
\begin{equation}
2\int_S^\infty dx \frac{1}{\sqrt{2\pi}}e^{-x^2/2} 
=
\int_{z_0(S)}^\infty dz \frac{z^{n/2-1}e^{-z/2}}{2^{n/2}\Gamma\left(n/2\right)}
\end{equation}
where $n$ is the number of bins in our LL construction and $z_0(S)$ is the LL necessary to give a significance of  $S$. 
If the data is treated with only one bin ($n=1$), then the LL approaches the significance squared $S=\sqrt{LL}$. 
We review the derivation of the LL from a Poisson probability in Appendix \ref{app:LL}.  

\begin{figure}
\centerline{\epsfig{file=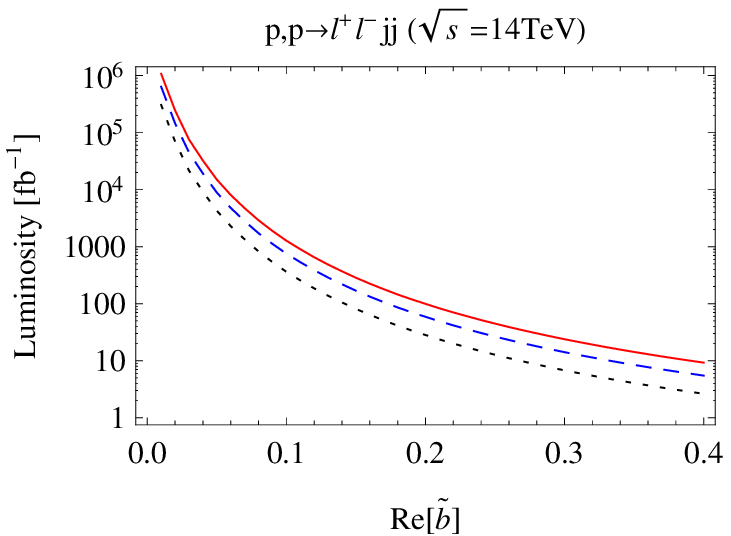,width=8.2cm}~\epsfig{file=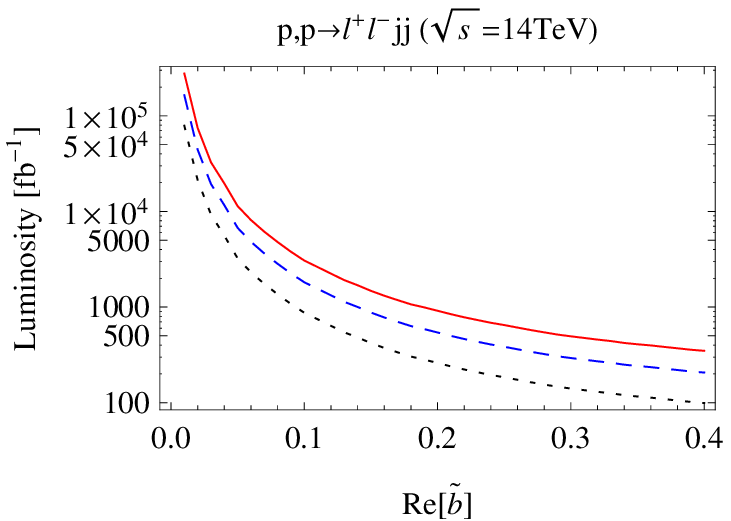,width=8.3cm}}
\caption{\label{fig:lLL}The integrated luminosity required to measure the absolute value of $\tilde b$ on the left panel,
and the sign of it  on the right panel at 1$\sigma$ (dotted line), 3$\sigma$ (dashed line) and 5$\sigma$ (solid line) level.}
\end{figure}

In our analysis, we used 14 bins so that a 1, 3 and 5$\sigma$ deviation corresponds with a LL of
\begin{eqnarray}
z_0(1)=15.9, \quad  z_0(3)=33.2,\quad  z_0(5)=56.0.
\label{zs}
\end{eqnarray}
Based on these results and matching the LL (fb) in Fig.~\ref{LLvsBt}(a), 
we can invert to obtain the integrated luminosity required to achieve a 1, 3 and 5$\sigma$ deviation of this observable. 
We  present the results in Fig.~\ref{fig:lLL}, 
for the integrated luminosity required to measure the absolute value of $\tilde b$ on the left panel
at 1$\sigma$ (dotted line), 3$\sigma$ (dashed line) and 5$\sigma$ (solid line) level.
We find that for $\tilde{b}=0.25$, we can measure the absolute value of the asymmetry at 3$\sigma$ (5$\sigma$) with approximately 30 fb$^{-1}$ (50 fb$^{-1}$). 
To determine the sign of the asymmetry, we must determine the probability that the asymmetry with one sign could fluctuate to look like the asymmetry with the opposite sign.  Following a similar procedure, but using the distribution with $\tilde{b}$ taking the opposite sign for the expectation in each bin (the $\nu_i$), we present the integrated luminosity to establish the sign of the asymmetry at 1, 3 and 5$\sigma$ level in the right panel of Figure \ref{fig:lLL}. We find that it takes approximately 380 fb$^{-1}$ (650 fb$^{-1}$) to determine its sign at 3$\sigma$ (5$\sigma$) for $\tilde{b}=0.25$.

\section{Discussion}
\label{sec:discussion}

The $ZZH$ interaction in the context of possible CP violation has been extensively studied for $e^{+}e^{-}$ 
colliders \cite{Hagiwara:1993sw}. 
Due to the neutrality of the initial state and well-constrained kinematics, it is found that an  $e^{+}e^{-}$ linear
collider could have significant sensitivity to probe the CP-odd coupling of $ZZH$, especially if beam 
polarization is achievable. Our work first established the feasibility to test the CP violation effects via similar
genuine CP-odd variables particularly suitable for the LHC for a light Higgs boson well below the $ZZ$ threshold. 

Assuming the observation of a light Higgs boson at the LHC, we have demonstrated the extent to which a CP-violating interaction in the
$ZZH$ vertex could be explored. 
It has been shown recently that this channel may become one of the Higgs discovery channels due to the improved techniques of studying jet substructure \cite{Butterworth:2008iy} and superstructure \cite{Gallicchio:2010sw}, and it is thus conceivable that some further
%since our major focus is on the CP asymmetry, it is beyond the scope of this paper to go through these newly developed techniques to 
%
enhancement for the signal-to-background ratio may be achieved with more sophisticated kinematical considerations. 
The LHC reach for discovering the CP-violation in Higgs couplings reported in this paper should be taken as a conservative estimate.
 Furthermore, our proposal is applicable
to any Higgs mass, as long as the production rate is sizable and Higgs
decay is identifiable. 

Besides the CP studies for a heavy Higgs boson \cite{Chang:1992tu}, 
the coupling $VVH$ for a light Higgs was also explored via 
%If not relying on genuine CP-odd observables, 
the weak boson fusion (WBF) production channel 
%can be explored to probe the Higgs CP property 
\cite{Plehn:2001nj}. It was found that CP-even and CP-odd operators may lead to qualitatively different angular distributions.
However, due to the lack of particle charge identification, no CP-odd observables can be constructed for the WBF processes. 

In this article, we have only studied the effect of $\bt$ in Eq.~(\ref{eq:operator}) beyond the tree-level SM. In principle, there could also be contributions to $a$ and $b$ which would contribute symmetrically to the signal.  If this is the case, their fluctuations could mimic small asymmetries requiring a more detailed LL analysis.  However, the required luminosity for discovery presented 
in Fig.~\ref{fig:Gauss} would still hold.

% ---------------------------------------------------------------------------------------------------------------------------------
\section{Conclusion}
\label{sec:conclusion} 

The need for new CP-violating interactions  to explain the observed matter-antimatter asymmetry
is pressing, and the unexplored Higgs sector may hold the key.  
If the LHC discovers the Higgs boson(s),
% it is our hope that the LHC will discover the Higgs boson(s), 
a detailed study of the properties of the Higgs will then begin. 
If the Higgs boson is heavy enough to decay to $W^+W^-,\ ZZ$ or $t\bar t$,
then it is quite feasible to construct CP-odd observables to test properties of the interactions.
However, if the Higgs boson is light and mainly decays to light fermion pairs,
then it would be extremely challenging to test the couplings $VVH$ or $H f\bar f$.

We studied two genuine CP-odd observables at the LHC to test the CP
property of the $ZZH$ interaction for a Higgs boson with mass below the threshold
to a pair of gauge bosons via the process $pp\to ZH \to \ell^{+} \ell^{-}\  b\bar b$. 
We showed that including a CP-odd $ZZH$ coupling and after selective kinematical cuts to suppress 
the SM backgrounds, we are able to extract the CP-asymmetries in the signal events. 
With a CP violating coupling $\tilde{b}\simeq0.25$, a CP asymmetry may be established at a
3$\sigma$ (5$\sigma$) level with an integrated luminosity of 30
(50) $\fbi$ at the LHC.

\vskip 0.2cm \noindent

%\acknowledgments
{\it Acknowledgments:} We thank G. Valencia for helpful discussions.  
T.H.~was  supported in part by the 
U.S.~Department of Energy under grant No.~DE-FG02-95ER40896.
Y.L.~was supported by the US DOE under contract DE-FG02-08ER41531.
N.D.C.~ was supported by the US NSF under grant number PHY-0705682.

%%%%%%%%%%%%%%%%%%%%%%%%%%%%%%%%%%%%%%%%%%%%%%%
%
%\title{How the log likelihood compares to Equation (5) in CMS Note 2005/004.}
%\author{Neil Christensen}
%\begin{document}
%\maketitle

%\appendix{The Log Likelihood}
\appendix
\section{\label{app:LL}The Log Likelihood}

The poisson probability of finding $n$ events when $\nu$ are expected is
\begin{equation}
f(n,\nu) = \frac{\nu^ne^{-\nu}}{n!} .
\end{equation}
The likelihood for finding $n_i$ events where $\nu_i$ are expected in $N$ bins is given by
\begin{equation}
L \propto \prod_i \frac{\nu_i^{n_i}e^{-\nu_i}}{n_i!} .
\end{equation}
To get a normalized likelihood, we divide it by the same form where $\nu_i=n_i$
\begin{equation}
L = \frac{
\prod_i \frac{\nu_i^{n_i}e^{-\nu_i}}{n_i!}
}{
\prod_i \frac{n_i^{n_i}e^{-n_i}}{n_i!}
}
=
\prod_i\left(\frac{\nu_i}{n_i}\right)^{n_i} e^{n_i-\nu_i} .
\end{equation}
The ``log likelihood'' is then defined as
\begin{equation}
LL = \mbox{ln}{1\over L^{2}} = 2\sum_i\left[n_i\mbox{ln}\left(\frac{n_i}{\nu_i}\right)+\nu_i-n_i\right].
\end{equation}
%
%This is the log likelihood.  It is important 
In the limit of a large number of events, the LL approaches a $\chi^2$ distribution
\begin{equation}
\chi^2(z,n) = \frac{z^{n/2-1}e^{-z/2}}{2^{n/2}\Gamma(n/2)}
\end{equation}
where $n$ is the number of bins $N$.  (If we were fitting $m$ parameters, then n would be $N-m$.)  The probability that the expectation value of $n$ would fluctuate up to at least $z_0$ or more is given by
\begin{equation}
P(z \geq z_0) = \int_{z_0}^\infty dz \chi^2(z,n)
\end{equation}
In order to determine the $z_0$ corresponding to a Gaussian significance of $S$, we simply set the probabilities of fluctuation equal as in
\begin{equation}
P(|x-S| \geq 0) = 2 \int_S^\infty dx \frac{1}{\sqrt{2\pi}}e^{-x^2/2} = P(z \geq z_0(S)) =  \int_{z_0(S)}^\infty dz \chi^2(z,n)
\end{equation}
and solve for $z_0(S)$.

%Since in the limit of large numbers of events it follows a $\chi^2$ distribution,
% given by 
%\begin{equation}
%f(z,n) = \frac{z^{n/2-1}e^{-z/2}}{2^{n/2}\Gamma(n/2)}
%\end{equation}
%where $n$ is the number of degrees of freedom (number of bins in our case).  If $n=1$ then $5\sigma$ corresponds with  $\chi^2=5^2=25$.
%this  $LL$ is related (or interpreted) to the statistical significance ...

%Now, if there is only one bin, this is:
%\begin{equation}
%LL = 2\left[n\mbox{ln}\left(\frac{n}{\nu}\right)+\nu-n\right]
%\end{equation}
%Since there is only one bin, this is equal to the significance squared.  So, the significance is the square root:
%\begin{equation}
%S = \sqrt{2\left[n\mbox{ln}\left(\frac{n}{\nu}\right)+\nu-n\right]}
%\end{equation}
%Finally, the number of expected events is the background ($\nu=b$) and the number obtained is the background plus signal ($n=b+s$) 
%giving:
%\begin{equation}
%S = \sqrt{2\left[(s+b)\mbox{ln}\left(1+\frac{s}{b}\right)-s\right]}
%\end{equation}
%This is the formula that  in CMS Note 2005/04.
%

%%%%%%%%%%%%%%%%%%%%%%%%%%%%%%%%%%%%%%%%%%%%%%%

\end{document}